\documentclass[
 reprint,
 amsmath,amssymb,
 aps,prl
]{revtex4-2}

\usepackage{graphicx}
\usepackage{dcolumn}
\usepackage{bm}
\usepackage[hidelinks]{hyperref}

\newcommand{\dr}{\mathrm{d}}
\newcommand{\be}{\begin{equation}}
\newcommand{\ee}{\end{equation}}
\newcommand{\bea}{\begin{eqnarray}}
\newcommand{\eea}{\end{eqnarray}}
\newcommand{\RR}{\mathbb R}
\newcommand{\pd}{\partial}

\begin{document}

\preprint{}




\title{Topological defects as lagrangian correspondences}

\author{Alex S.~Arvanitakis\\\small \texttt{alex.s.arvanitakis@vub.be}}
\affiliation{%
Theoretische Natuurkunde, Vrije Universiteit Brussel\nobreak, and the International Solvay
Institutes,\\
Pleinlaan 2, B-1050 Brussels, Belgium
}%

\date{\today}

\begin{abstract}
Topological defects attract much recent interest in high-energy and condensed matter physics because they encode (non-invertible) symmetries and dualities. We study codimension-1 topological defects from a hamiltonian point of view, with the defect location playing the role of `time'. We show that the Weinstein symplectic category governs topological defects and their fusion: each defect is a lagrangian correspondence, and defect fusion is their geometric composition. We illustrate the utility of these ideas by constructing S- and T-duality defects in string theory, including a novel topology-changing non-abelian T-duality defect.
\end{abstract}

\maketitle

Quantum field theories are about more than the physics of pointlike particles. Even outside of String/M-theory --- whose \emph{raison d'\^etre} is the physics of fundamental extended objects --- one finds that line, surface, and higher-dimensional objects have much to say, even if they were not originally `put in by hand' in the theory in question. Early examples are Wilson loops in QCD; their VEVs probe quark confinement.

Extended objects and operators supported on such appear generically as mediators of dualities, global symmetries, and their `higher'/`generalised' counterparts \cite{Gaiotto:2014kfa}.  In this paper we will be exlusively study codimension-1 extended objects (of dimension $\dim M-1$ if the theory in question lives on the manifold $M$) which we will call \emph{defects}. (They are also called `walls' or `interfaces'.)

We will in particular be interested in \emph{topological defects} --- ones which may be freely deformed such that correlation functions are unchanged --- due to their close connection to dualities \cite{Fuchs:2007fw,Sarkissian:2008dq,Gaiotto:2008ak,Kapustin:2009av}. The physical picture is simple: we envision a scenario where the defect locus separates spacetime $M$ into phase 1 (say inside) and phase 2 (say outside), where phases 1 and 2 can be two copies of the same theory, or possibly two different theories. For e.g.~a spherical defect of radius $t$, as $t$ goes from $0$ to $+\infty$ we observe a transition from phase 2  to phase 1; by inserting operators inside or outside this defect sphere, we thus obtain the action of the associated duality transformation on all local operators. Moreover, two topological defects can be \emph{fused} by deforming them so they lie close to each other; this connects the algebraic aspects of symmetry --- (semi-)group structure --- to the defect picture. Fusion accommodates both conventional symmetries --- related to defects that may fuse into the ``invisible'' defect --- as well as more exotic non-invertible symmetries.

Since the presence of one or more defects breaks Lorentz invariance, we might as well study them in a \emph{hamiltonian formulation}. The hamiltonian `time' $t$ will be such that $t=\text{const.}$ is the defect locus; varying the constant gives a family of defects. We will discuss arbitrary systems in a hamiltonian path-integral formulation, working in finite dimensions for technical simplicity.

The main take-away from this work will be that \textbf{topological defects and their fusion furnish the Weinstein symplectic category} \cite{weinstein1981symplectic,weinstein1982symplectic} of symplectic manifolds with lagrangian correspondences --- which are related to canonical transformations in hamiltonian mechanics --- as morphisms. We will demonstrate in the Discussion section that this gives a practical and efficient way to identify topological defects in diverse physical contexts and to study their fusion. In the rest of this letter we will explain what this statement means and give a brief but reasonably complete account of why it is true.



We start with textbook hamiltonian mechanics and canonical transformations. Consider a \emph{phase-space action principle} for e.g.~a particle on $\RR$:
\be
\label{eq:phasespaceaction}
S[x,p]=\int_{T_\mathrm{I}}^{T_{\mathrm{F}}} \dr t\;\{ p(t)\dot x(t) -h\big(x(t),p(t)\big)\}
\ee
where $\dot x=\dr x/\dr t$. The hamiltonian $h$ is a function depending on a point $(x,p)\in \RR^2$ of \emph{phase space}; it characterises the system in question. (We assume it is time-independent for simplicity.) Time evolution with fixed initial and final boundary conditions $x(T_\mathrm{I})=x_\mathrm{I},x(T_\mathrm{F})=x_\mathrm{F}$ is determined by extremising $S$, yielding Hamilton's equations.

A \emph{canonical transformation} is supposed to express $x,p$ in terms of new variables $x_2,p_2$  such that the equations of motion arise from an action $S[x_2,p_2]$ of the form \eqref{eq:phasespaceaction}, possibly with different hamiltonian $h\to h_2$. This will be the case if there exists a function $F$ with
\be
p\dot x -h=p_2 \dot x_2 - h_2 +\dr  F/\dr t\,.
\ee
$F$ 
is called the \emph{generating function}.

With some assumptions, $F$ can determine the transformation. For example, assume the map $(x,p)\to(x_2,p_2)$ is such that we can unambiguously specify any point in phase space by fixing original and new position variables $(x,x_2)$, and also assume $F$ depends on $(x,x_2)$. (This is a ``type 1 generating function''.) the velocities $\dot x,\dot x_2$ are linearly independent of each other and of $(x,x_2)$, yielding
\be \label{eq:type1cantransf} p= \pd F/\pd x\,,\quad p_2=-\pd F/\pd x_2\,,\quad h_2=h;\ee
this can be solved to produce a map $(x,p)\to (x_2,p_2)$ under favourable conditions. (We return to this later.)


\section{Endpoint contributions for mechanics on manifolds}
Let us review a subtlety in hamiltonian dynamics on a symplectic manifold $M$, relevant whenever the symplectic form $\omega$ does not admit a  symplectic potential 1-form $\theta$ with $\omega=\dr \theta$ that is non-singular everywhere. (I.e.~$\omega$ is not exact. Here and henceforth we employ the language of differential forms; $\dr$ is the de Rham differential.) We assume $\omega$ obeys a Dirac quantisation condition,
\be
\label{eq:quantizationcondition}
\int_{C}\omega=2\pi \mathbb Z
\ee
for $C$ any closed 2-dimensional cycle inside $M$.

The previous section describes the case $M=T^\star \RR$ with symplectic form $\omega=\dr (p \dr x)=\dr p\wedge \dr x$, and $\theta=p\dr x$.  The problem of generalising the action \eqref{eq:phasespaceaction} to a symplectic manifold where $\omega $ is not exact is mathematically the same as the the problem of coupling a charged particle to the electromagnetic field generated by a magnetic monopole. The subtlety in the monopole case is also  that the electromagnetic potential $\mathcal A$ is not well-defined everywhere, so that the usual coupling $\int \dr t \;\mathcal A_\mu(x(t))\dot x^\mu$ to the particle worldline needs a careful prescription.

The correct prescription was given long ago by Wu and Yang \cite{wu1976dirac}. First, we cover $M$ by  open subsets $U_\alpha$ such that each $U_\alpha$ and all intersections $U_\alpha \cap U_\beta$ are contractible. Then for each $\alpha$ there exists a locally-defined symplectic potential 1-form $\theta_\alpha$ such that $\dr \theta_\alpha=\omega|_{U_\alpha}$ due to the Poincar\'e lemma. ($\omega|_{U_\alpha}$ is the restriction.)
On overlaps, using the Poincar\'e lemma again, we find scalar functions $g_{\alpha\beta}$ describing a `gauge transformation' from $\theta_\alpha$ to $\theta_\beta$:
\be
\theta_\beta-\theta_\alpha=g_{\alpha\beta}\,.
\ee
Given any curve $\gamma:[T_\mathrm{I},T_{\mathrm{F}}]\to M$, we write the analogue of the action \eqref{eq:phasespaceaction} by splitting the curve at arbitrary points within any overlap $U_\alpha \cap U_\beta$. If e.g.~the curve happens to lie entirely within two contractible patches, $U_\alpha\ni \gamma(T_\mathrm{I})$ and $U_\beta\ni \gamma(T_\mathrm{F})$, with $\gamma(t')\in U_\alpha\cap U_\beta$ then we write
\be
\label{eq:wuyangexample1overlap}
\int_{T_\mathrm{I}}^{t'} \gamma^\star\theta_\alpha +\int_{t'}^{T_\mathrm{F}}\gamma^\star\theta_\beta +g_{\alpha\beta}(\gamma(t'))\,.
\ee
The $g_{\alpha\beta}$ term ensures independence from the arbitrary choice of intermediate $t'$. 
Given the quantisation condition \eqref{eq:quantizationcondition}, the construction is independent of choices \cite{alvarez1985topological}. 

 The upshot is: we may make arbitrary choices of local symplectic potential, \emph{at the price of generating contributions that depend on the endpoints $\gamma(T_\mathrm{I}),\gamma(T_\mathrm{F})$} (and possibly on the intermediate points within overlaps). 
We thus write the action for hamiltonian mechanics on a symplectic manifold as (where we reinstated the hamiltonian $h\in C^\infty(M)$, and where $\gamma(T_\mathrm{I})\equiv p_\mathrm{I}$ $\gamma(T_\mathrm{F})\equiv p_\mathrm{F}$)
\bea
\label{eq:nodefectaction}
S[\gamma]= S_\mathrm{I}(p_\mathrm{I}) + S_\mathrm{F}(p_\mathrm{F})+\int_{T_\mathrm{I}}^{T_\mathrm{F}}\gamma^\star \theta - \gamma^\star(h)\dr t 
\eea
The ``endpoint contribution'' functions $S_\mathrm{I}$ and $S_\mathrm{F}$ compensate for the ambiguity in the choice of symplectic potential near the endpoints $p_\mathrm{I}$ and $p_\mathrm{F}$, while $\int_{T_\mathrm{I}}^{T_\mathrm{F}}\gamma^\star \theta$ is interpreted via the Wu-Yang prescription (as in \eqref{eq:wuyangexample1overlap}). In particular we can set $S_\mathrm{I}=S_\mathrm{F}=0$ at the price of fixing the symplectic potentials near each endpoint.


\section{Lagrangian correspondences and their composition}
A \emph{lagrangian submanifold}, or just \emph{lagrangian} $L$ of a symplectic manifold $M$ is a maximal submanifold where $\omega$ vanishes; explicitly, $\iota_L^\star\omega=0$ for $\iota_L$ the inclusion map $\iota_L:L\hookrightarrow M$. (In other words, $\omega(v_1,v_2)=0$ for all vectors $v_1,v_2$ tangent to $L$.) Maximality in the case $\dim M<\infty$ means $\dim L=\dim M/2$. For the particle on the line \eqref{eq:phasespaceaction}, both submanifolds $p=0$ and $x=0$ are lagrangian.

In general a choice of lagrangian submanifold amounts to a (local) choice of `position' and `momentum' variables on \emph{any} symplectic manifold: this is realised via the \emph{Weinstein lagrangian neighbourhood theorem}, which states that an open neighbourhood of $L$ in $M$ is isomorphic to an open neighbourhood of the zero section of the cotangent bundle $T^\star L$, and $L\subset M$ maps to the zero section $L\hookrightarrow T^\star L$. A crucial corollary is that given a lagrangian $L$, we can always find a coordinate system on $M$ with $\dim M/2$ `momenta' $p_a$ and `positions' $x^a$, and an `adapted' local symplectic potential $\theta_L$ with
\be
\label{eq:sympPotLagneigh}
\theta_L=p_a\dr x^a 
\ee
with $p_a=0$ for all $a=1,2,\cdots (\dim M/2)$ specifying the chosen lagrangian $L$. 

\emph{Canonical transformations are also properly understood as lagrangians}: for example, the ``type 1 transformation'' of \eqref{eq:type1cantransf} may be described via the lagrangian submanifold $L_0$ specified as the locus $p=p_2=0$ in the symplectic manifold $T^\star \RR\times T^\star \RR$, with symplectic form
\be
\dr p\wedge \dr x-\dr p_2\wedge \dr x_2.
\ee
The adapted potential is $\theta_{L_0}=p\dr x - p_2\dr x_2$. Then, given any function $F\in C^\infty(L_0)$, we find a  lagrangian $L_{F}$ specified by \eqref{eq:type1cantransf}; geometrically, $L_F$ is the image of $L_0$ under the hamiltonian flow generated by $F$. Conversely: the general apparatus of generating functions amounts to finding lagrangians specified, in this way, by functions $F$. (The ``type'' is encoded in the choice of $L_0$.) From this perspective the invertibility of \eqref{eq:type1cantransf} is immaterial.

A \textbf{lagrangian correspondence} from symplectic manifold $(M_1,\omega_1)$ to symplectic manifold $(M_2,\omega_2)$ is a lagrangian $L_{12}$ inside $(M_1\times M_2,\omega_2-\omega_1)$. We have just seen how lagrangian correspondences appear as canonical transformations in mechanics. The concept appears more prominently in mathematics in the context of the \emph{Weinstein symplectic category} \cite{weinstein1981symplectic,weinstein1982symplectic}. Put briefly, the idea is that lagrangian correspondences should be considered as morphisms between symplectic manifolds. The motivation  comes from quantisation: we should assign a Hilbert space $H_1$ to a symplectic manifold $M_1$, and --- perhaps less obviously --- a state $|\psi\rangle$ to a lagrangian $L\subset M$. If $M_1\times M_2$ is assigned the tensor product $H_1\otimes H_2$, then a linear map $H_1\to H_2$ should map to elements of $H_1^\star\otimes H_2$, which should come from  lagrangian correspondences $L_{12}$.

Therefore, two lagrangian correspondences $L_{12}$ (from $M_1$ to $M_2$) and $L_{23}$ (from $M_2$ to $M_3$) ought to be composable into a third one $L_{13}$. We can define the set
\be
\label{eq:lagcorrcomp}
L_{13}=\Pi_{13}\Big((L_{12}\times L_{23})\cap (M_1\times \Delta_{M_2}\times M_3)\Big)
\ee
where $\Delta_{M_2}:M_2\to M_2\times M_2$ is the diagonal, and $\Pi_{13}:M_1\times M_2\times M_2\times M_3\to M_1\times M_3$ is the projection. This set is not  a manifold \emph{unless} the intersection is transverse; in that case $L_{13}$ is an immersed submanifold and $L_{13}$ is a lagrangian correspondence called the \textbf{geometric composition} of $L_{12}$ and $L_{23}$. We refer to \cite[Section 2]{wehrheim2010quilted} for more on lagrangian correspondences, and to the recent review \cite{abouzaid2022functoriality} for context and mathematical applications thereof.

\section{Topological hamiltonian defects}
We now study a defect on the worldline of a particle at time $t=t_{12}$ via a generalisation of the action \eqref{eq:nodefectaction}. We write $S=I_{12}+S^\mathrm{D}_{12}$, with $S^\mathrm{D}_{12}(p_{12})$ a function depending on the defect location $p_{12}\equiv(\gamma_1(t_{12}),\gamma_2(t_{12}))$, and
\bea
I_{12}=\int_{-\infty}^{t_{12}}\gamma_1^\star \theta_1 - \gamma_1^\star(h_1)\dr t +\int_{t_{12}}^{+\infty}\gamma_2^\star \theta_2 - \gamma_2^\star(h_2)\dr t\,.
\eea

The defect separates ``phase 1'' $(t<t_{12})$ where the particle moves on the symplectic manifold $(M_1,\omega_1)$ with hamiltonian $h_1$ along the curve $\gamma_1:(-\infty,t_{12}]\to M_1$ from ``phase 2'' $(t>t_{12})$. Henceforth we are not concerned with the endpoints $T_\mathrm{I},T_\mathrm{F}$, so we set $T_\mathrm{I}=-\infty$, $T_\mathrm{F}=+\infty$. However, the \emph{defect action} $S^\mathrm{D}_{12}$ will be important.

A \textbf{topological defect} is one that can be moved around at no cost, whence the defining condition is
\be
\label{eq:topdefectconditionworldline}
\frac{\dr S}{\dr t_{12}}=0\,.
\ee
We will derive constraints on $p_{12}$ from this.

To proceed we employ a \emph{folding trick} \cite{wong1994tunneling}: define the ``folded'' curve $\Gamma_{12}$ on $M_1\times M_2$ as
\be
\label{eq:foldedcurve}
\Gamma_{12}(\tau)=\big(\gamma_1(t_{12}-\tau),\gamma_2(t_{12}+\tau)\big)\,,
\ee
then $S$ takes the form \eqref{eq:nodefectaction} with a single endpoint at $\tau=0$:
\be
\label{eq:oncefoldedaction}
S[\Gamma_{12}]=S^\mathrm{D}_{12}(p_{12})+ \int_{0}^{+\infty}\Gamma_{12}^\star \Theta_{12} - \Gamma_{12}^\star(H_{12})\dr \tau\,.
\ee
Here $\Theta_{12}=\theta_2-\theta_1$ is a (local) symplectic potential for the symplectic form $\omega_2-\omega_1$ on $M_1\times M_2$. (There was a sign change because $t=t_{12}-\tau$ is orientation-reversing.) The integral is again defined via the Wu-Yang prescription, which is sensible since $\omega_2-\omega_1$ satisfies the quantisation condition \eqref{eq:quantizationcondition} if $\omega_1$ and $\omega_2$ do. The ``folded'' hamiltonian is the sum $H_{12}=h_1+h_2$.

Consistency of the variational principle \eqref{eq:oncefoldedaction} requires boundary conditions at $\tau=0$. We anticipate a future calculation and select the boundary condition $\Gamma_{12}(0)\equiv p_{12}\in L_{12}$ for $L_{12}$ a lagrangian correspondence from $M_1$ to $M_2$, i.e.~a lagrangian for $(M_1\times M_2,\omega_2-\omega_1)$. We can thus employ the lagrangian neighbourhood theorem and produce an adapted (to $L_{12}$) potential $\Theta_{12}=P_A \wedge \dr X^A$ near $\Gamma_{12}(0)=p_{12}$, so that the boundary condition is simply $P_A(0)=0$ for \emph{all} curves. (The index $A$ takes $(\dim M_1+\dim M_2)/2$ values.) Since we have made a choice of symplectic potential near the endpoint, off-shell, we can thus set $S^\mathrm{D}_{12}=0$.



If we write $\dr/\dr t_{12}=\delta$, then using \eqref{eq:foldedcurve}
\be
\delta \Gamma_{12}^\star(H_{12})=\frac{\dr}{\dr \tau}\Gamma_{12}^\star (h_2-h_1)\,,
\ee
and, (where $\Omega_{12}=\omega_2-\omega_1$)
\be
\label{eq:deltaTheta12}
\delta (\Gamma_{12}^\star \Theta_{12})=\frac{\dr}{\dr \tau}(P_A\delta X^A)\dr \tau +\Omega_{12}(\delta \Gamma_{12},\dot \Gamma_{12})\dr \tau\,.
\ee
In the last term, $\dot \Gamma_{12}$ is the tangent vector to the curve \eqref{eq:foldedcurve}, while $\delta \Gamma_{12}$ is the vector obtained by the $\delta=\dr/\dr t_{12}$ derivative. However, the two are related by $\delta \Gamma_{12}=C \dot \Gamma_{12}$ for a certain constant matrix $C$ (due to \eqref{eq:foldedcurve}); $C$ satisfies
\be
C^2=1\,,\qquad \Omega_{12}(C V,C U)=\Omega_{12}(V,U)\,,
\ee
whence $\Omega_{12}(\delta \Gamma_{12},\dot \Gamma_{12})=0$. (In ``factorised'' coordinates, $C$ changes the sign of vectors pointing along $M_1$, from which these identities are now obvious.) Using the boundary condition $P_A=0$ at $\tau=0$, the first term in \eqref{eq:deltaTheta12} also fails to contribute, so we obtain
\be
\delta S=\Gamma_{12}^\star(h_1-h_2)|_{\tau=0}=0.
\ee

Therefore, the topological defect condition \eqref{eq:topdefectconditionworldline} holds \emph{if} the defect location $p_{12}$ lies on a lagrangian correspondence $L_{12}$, \emph{and} the hamiltonians $h_{1,2}$ match on each side of the defect. This is essentially the matching of stress-energy tensors on topological defects in field theory (see e.g.~\cite{Kapustin:2009av}).



\bigskip
We now discuss fusion. We take two topological defects at locations $t=t_{12}$ and $t=t_{23}$ that define a segmented worldline along symplectic manifolds $M_{1}$ (for $t\leq t_{12}$), $M_2$ (for $t_{12}\leq t \leq t_{23}$) and $M_3$ (for $t\geq t_{23}$), with symplectic forms and hamiltonian functions $(\omega_1,h_1)$ on $M_1$, etc. We have associated lagrangian correspondences $L_{12}$ (to the $t_{12}$ defect) and $L_{23}$ (to the $t_{23}$ defect) as before.

To write the action we introduce the worldline 1-forms
\be
\alpha_{1,2,3}=\gamma_{1,2,3}^\star \theta_{1,2,3} - \gamma_{1,2,3}^\star(h_{1,2,3})\dr t\,.
\ee
For the action we take simply
\be
S=\int_{-\infty}^{t_{12}} \alpha_1  +\int^{t_{23}}_{t_{12}} \alpha_2 + \int_{t_{23}}^{+\infty} \alpha_3\ee
which is again interpreted via the Wu-Yang prescription and where we have made a choice of symplectic potentials close to $t_{12}$ and $t_{23}$ that are adapted to the lagrangian correspondences so as to set the defect actions to zero.

Since both defects are topological, we can shift them around freely. By \emph{defect fusion}, we mean the defect obtained in the coincidence limit $t_{12}\to t_{23}$. This ought to be a topological defect between $M_1$ and $M_3$, thus a lagrangian correspondence $L_{13}$ from $M_1$ to $M_3$.

We  will calculate $L_{13}$ from $L_{12}$ and $L_{23}$. Let us write
\be
\label{eq:fusiontrick} S=\int_{-\infty}^{t_{12}} \alpha_1 +\int_{t_{12}}^{+\infty} \alpha_2 +\int_{-\infty}^{t_{23}} \alpha_2  +\int_{t_{23}}^{+\infty} \alpha_3 -\int_{-\infty}^{+\infty} \alpha_2\,. \ee
This entails an arbitrary extension of $\gamma_2$ to the entire real line; the arbitrariness will drop out shortly. The point of the rewriting is that we may `fold' the \emph{first four} terms pairwise to obtain terms of the form \eqref{eq:oncefoldedaction}. This yields
\be
\int_{0}^{+\infty}\Gamma_{12}^\star \Theta_{12}+\Gamma_{23}^\star \Theta_{23} - (\Gamma_{12}^\star(H_{12})+\Gamma_{23}^\star(H_{23}))\dr \tau\,.
\ee
Here $\Theta_{12}$ is a symplectic potential for $\omega_2-\omega_1$, $\Theta_{23}$ is one for $\omega_3-\omega_2$, and $H_{12}=h_1+h_2,H_{23}=h_2+h_3$. $\Gamma_{12}$ is the curve \eqref{eq:foldedcurve}, while $\Gamma_{23}$ is given similarly.

We interpret this via the twice-folded curve $\Gamma_{1223}(\tau)=(\Gamma_{12}(\tau),\Gamma_{23}(\tau))$ on the manifold $M_1\times M_2\times M_2\times M_3$:
\be
\big(\gamma_1(t_{12}-\tau),\gamma_2(t_{12}+\tau),\gamma_2(t_{23}-\tau),\gamma_3(t_{23}+\tau)\big)\,.
\ee
For $\tau=0$ this is $(p_{12},p_{23})$ where $p_{12}\in L_{12}$ and $p_{23}\in L_{23}$.

 Since both original defects are topological, we can send e.g.~$t_{23}\to 0$, then the coincidence limit is $t_{12}\to 0$, and the twice-folded curve becomes
\be
\label{eq:twicefolded}
\Gamma_{1223}(\tau)=\big(\gamma_1(-\tau),\gamma_2(+\tau),\gamma_2(-\tau),\gamma_3(+\tau)\big)\,.
\ee
We see $\gamma_2$ is traversed twice as $\tau$ goes from $0$ to $+\infty$; this produces an integral that cancels the last term of \eqref{eq:fusiontrick}, thus eliminating the `intermediate' manifold $M_2$. Moreover, for $\tau=0$ we see \eqref{eq:twicefolded} takes the form $(p_1,p_2,p_2,p_3)$ for points $p_{1,2,3}\in M_{1,2,3}$. Since $\Gamma_{2}(0)\equiv p_{12}=(p_1,p_2)$ and $\Gamma_{23}(0)\equiv p_{23}=(p_2,p_3)$ each lie on $L_{12}$ and $L_{23}$, we have arrived precisely at the geometric composition of the lagrangian correspondences \eqref{eq:lagcorrcomp}. The hamiltonians on each side of the fused defect $L_{13}$ trivially agree with each other, so this completes the argument.

\section{Discussion}
Since canonical transformations (in the physics sense) give rise to lagrangian correspondences, we can construct a topological defect for every canonical transformation. We exploit this to construct duality defects. As a first example, take $d=4$ $\mathrm{U}(1)$ euclidean gauge theory, with complex coupling $\tau=\theta/(2\pi)+4\pi i/g^2$ (of which $g^{-2}$ is the coupling and $\theta$ the theta angle). The phase space for electromagnetism is spanned by the `positions' $\mathcal A_i(\sigma)$ ($i=1,2,3$, $\sigma \in \RR^3$) and their conjugate `momenta' $\Pi^i(\sigma)$; the former are the spatial components of the gauge potential 1-form $\mathcal A$. In \cite{Lozano:1995aq} there is a canonical transformation that implements $\tau\to -\tau^{-1}$;  thus, there exists a topological defect between the theory with coupling $\tau$ and the theory with coupling $-\tau^{-1}$, as was found in \cite{Gaiotto:2008ak,Kapustin:2009av} entirely differently. Explicitly: we have the type 1 generating function 
\be
F=\int_{M_4}\mathcal F\wedge \tilde{\mathcal F}
\ee
in terms of the original and dual field strengths $\mathcal F=\dr \mathcal A,\tilde{\mathcal F}=\dr \tilde{\mathcal A}$. (For this to work we need the \emph{reduced} phase space for electromagnetism, obtained by symplectic reduction modulo the Gauss law constraint $\pd_i \Pi^i=0$.)  We can similarly obtain S-duality defects on the worldvolume of the D3 brane in type IIB string theory \cite{Lozano:1997cy} (including fermions \cite{igarashi1998self}), as well as between a IIB superstring and a D1-brane \cite{igarashi1998canonical}.


We may also easily recover worldsheet T-duality defects from the T-duality canonical transformation given in \cite{alvarez1994canonical}. This extends to other flavours of T-duality, including Poisson-Lie (via \cite{Klimcik:1995dy,Sfetsos:1996xj}) and even fermionic T-duality (via \cite{Sfetsos:2010xa}); for the latter case topological defects were also found in \cite{Elitzur:2013ut} with different methods. In the Poisson-Lie case, we have thus made contact with the recent ``Poisson-Lie defects'' of \cite{Demulder:2022nlz}.

We can say more about Poisson-Lie T-duality defects, however. In \cite{Arvanitakis:2021lwo} we recently co-introduced a joint generalisation of topological \cite{bouwknegt2004t,bouwknegt2004s} and Poisson-Lie T-dualities in the form of lagrangian correspondences between the phase spaces for string propagation on a principal \emph{bibundle} $G\hookrightarrow M\dashrightarrow B$ and its dual bibundle $\tilde G\hookrightarrow \tilde M\dashrightarrow B$, whose fibres $G$ and $\tilde G$ are Poisson-Lie dual groups; this `bibundle duality' can realise topology changes in the global fibration structure of target space akin to those of topological (abelian) T-duality, which is a special case. With the result of the current paper we can thus realise bibundle duality via a topological worldsheet defect. (This way, the Drinfeld double bibundles of \cite{Arvanitakis:2021lwo} give novel `bibranes' in the terminology of \cite{Fuchs:2007fw}.)

There are also examples of dualities that may be understood this way outside of string theory (which is admittedly over-represented above on account of the author's preferences/expertise). For instance, bosonisation in two-dimensional field theory has been described in terms of canonical transformations \cite{Ikehashi:1993np,Bordner:1997pv} and thus may be realised via topological defects, like S-duality in $d=4$ gauge theory was above.

There are a few notable omissions from our brief treatment. The first is the introduction of time-dependence (in the hamiltonians $h_1,h_2$) which \emph{a priori} motivates introducing time-dependence in the defect action $S^\mathrm{D}_{12}(p_{12},t_{12})$. This changes little: choosing $\Gamma_{12}(0)\in L_{12}$ as a boundary condition (for time-independent $L_{12}$) along with appropriate endpoint contributions eventually forces $S^\mathrm{D}_{12}=S^\mathrm{D}_{12}(t_{12})$ without loss of generality. The effect is to introduce a discontinuity in the hamiltonians, so $h_2=h_1+\pd S^\mathrm{D}_{12}/\pd t_{12}$ on the defect.

Another omission has to do with the transversality issue discussed below \eqref{eq:lagcorrcomp}: the fusion of two topological defects may fail to be a defect, in the sense that the corresponding lagrangian may be singular (as a manifold). (For this reason Weinstein himself called it the ``symplectic `category'''.) Fortunately, it was shown in \cite[Proposition 5.2.1]{wehrheim2010quilted}\cite[Theorem 2.3]{wehrheim2011quilted} that the intersection of \eqref{eq:lagcorrcomp} is rendered transverse if the lagrangians are perturbed appropriately; this could be realised in our picture by switching on small defect actions ($S^\mathrm{D}_{12}$ and $S^\mathrm{D}_{23}$). The physical implications of this (if any) are left for the future.

Finally, we will also be leaving the treatment of gauge theories, i.e.~hamiltonian systems with first-class constraints for the future. That case includes temporal reparameterisation-invariance, which introduces many subtleties; however, we note that composition of correspondences already allows us to construct topological defects in the \emph{reduced (gauged) theory} from topological defects in the original theory: the key point is that coisotropic reduction gives a lagrangian correspondence between the original and reduced symplectic manifolds \cite[Section 3]{weinstein2010symplectic}. In this context we also expect that introducing degrees of freedom lying on the defect might be necessary, as is common when discussing topological defects in field theory. 

\begin{acknowledgments}\paragraph{Acknowledgments:} This project arose from inspiring interactions with Lewis Cole, Saskia Demulder, and Dan Thompson, who also provided helpful feedback. I would also like to thank Jonny Evans for providing helpful references (on Mathoverflow \cite{436905}) and Nate Bottman for a clarification on his work.

I am supported by the FWO-Vlaanderen through the project G006119N, as well as by the Vrije Universiteit Brussel through the Strategic Research Program “High-Energy Physics”. I am also supported by an FWO Senior Postdoctoral Fellowship (number 1265122N).  I am grateful to the Mainz Institute for Theoretical Physics (MITP) of the DFG Cluster of Excellence PRISMA\textsuperscript{*} (Project ID 39083149), for its hospitality and its partial support during the completion of this work. 
\end{acknowledgments}

\bibliography{mybib}

\begin{thebibliography}{30}%
\makeatletter
\providecommand \@ifxundefined [1]{%
 \@ifx{#1\undefined}
}%
\providecommand \@ifnum [1]{%
 \ifnum #1\expandafter \@firstoftwo
 \else \expandafter \@secondoftwo
 \fi
}%
\providecommand \@ifx [1]{%
 \ifx #1\expandafter \@firstoftwo
 \else \expandafter \@secondoftwo
 \fi
}%
\providecommand \natexlab [1]{#1}%
\providecommand \enquote  [1]{``#1''}%
\providecommand \bibnamefont  [1]{#1}%
\providecommand \bibfnamefont [1]{#1}%
\providecommand \citenamefont [1]{#1}%
\providecommand \href@noop [0]{\@secondoftwo}%
\providecommand \href [0]{\begingroup \@sanitize@url \@href}%
\providecommand \@href[1]{\@@startlink{#1}\@@href}%
\providecommand \@@href[1]{\endgroup#1\@@endlink}%
\providecommand \@sanitize@url [0]{\catcode `\\12\catcode `\$12\catcode
  `\&12\catcode `\#12\catcode `\^12\catcode `\_12\catcode `\%12\relax}%
\providecommand \@@startlink[1]{}%
\providecommand \@@endlink[0]{}%
\providecommand \url  [0]{\begingroup\@sanitize@url \@url }%
\providecommand \@url [1]{\endgroup\@href {#1}{\urlprefix }}%
\providecommand \urlprefix  [0]{URL }%
\providecommand \Eprint [0]{\href }%
\providecommand \doibase [0]{https://doi.org/}%
\providecommand \selectlanguage [0]{\@gobble}%
\providecommand \bibinfo  [0]{\@secondoftwo}%
\providecommand \bibfield  [0]{\@secondoftwo}%
\providecommand \translation [1]{[#1]}%
\providecommand \BibitemOpen [0]{}%
\providecommand \bibitemStop [0]{}%
\providecommand \bibitemNoStop [0]{.\EOS\space}%
\providecommand \EOS [0]{\spacefactor3000\relax}%
\providecommand \BibitemShut  [1]{\csname bibitem#1\endcsname}%
\let\auto@bib@innerbib\@empty
\bibitem [{\citenamefont {Gaiotto}\ \emph {et~al.}(2015)\citenamefont
  {Gaiotto}, \citenamefont {Kapustin}, \citenamefont {Seiberg},\ and\
  \citenamefont {Willett}}]{Gaiotto:2014kfa}%
  \BibitemOpen
  \bibfield  {author} {\bibinfo {author} {\bibfnamefont {D.}~\bibnamefont
  {Gaiotto}}, \bibinfo {author} {\bibfnamefont {A.}~\bibnamefont {Kapustin}},
  \bibinfo {author} {\bibfnamefont {N.}~\bibnamefont {Seiberg}},\ and\ \bibinfo
  {author} {\bibfnamefont {B.}~\bibnamefont {Willett}},\ }\bibfield  {title}
  {\bibinfo {title} {{Generalized Global Symmetries}},\ }\href
  {https://doi.org/10.1007/JHEP02(2015)172} {\bibfield  {journal} {\bibinfo
  {journal} {JHEP}\ }\textbf {\bibinfo {volume} {02}},\ \bibinfo {pages}
  {172}},\ \Eprint {https://arxiv.org/abs/1412.5148} {arXiv:1412.5148 [hep-th]}
  \BibitemShut {NoStop}%
\bibitem [{\citenamefont {Fuchs}\ \emph {et~al.}(2008)\citenamefont {Fuchs},
  \citenamefont {Schweigert},\ and\ \citenamefont {Waldorf}}]{Fuchs:2007fw}%
  \BibitemOpen
  \bibfield  {author} {\bibinfo {author} {\bibfnamefont {J.}~\bibnamefont
  {Fuchs}}, \bibinfo {author} {\bibfnamefont {C.}~\bibnamefont {Schweigert}},\
  and\ \bibinfo {author} {\bibfnamefont {K.}~\bibnamefont {Waldorf}},\
  }\bibfield  {title} {\bibinfo {title} {{Bi-branes: Target space geometry for
  world sheet topological defects}},\ }\href
  {https://doi.org/10.1016/j.geomphys.2007.12.009} {\bibfield  {journal}
  {\bibinfo  {journal} {J. Geom. Phys.}\ }\textbf {\bibinfo {volume} {58}},\
  \bibinfo {pages} {576} (\bibinfo {year} {2008})},\ \Eprint
  {https://arxiv.org/abs/hep-th/0703145} {arXiv:hep-th/0703145} \BibitemShut
  {NoStop}%
\bibitem [{\citenamefont {Sarkissian}\ and\ \citenamefont
  {Schweigert}(2009)}]{Sarkissian:2008dq}%
  \BibitemOpen
  \bibfield  {author} {\bibinfo {author} {\bibfnamefont {G.}~\bibnamefont
  {Sarkissian}}\ and\ \bibinfo {author} {\bibfnamefont {C.}~\bibnamefont
  {Schweigert}},\ }\bibfield  {title} {\bibinfo {title} {{Some remarks on
  defects and T-duality}},\ }\href
  {https://doi.org/10.1016/j.nuclphysb.2009.04.016} {\bibfield  {journal}
  {\bibinfo  {journal} {Nucl. Phys. B}\ }\textbf {\bibinfo {volume} {819}},\
  \bibinfo {pages} {478} (\bibinfo {year} {2009})},\ \Eprint
  {https://arxiv.org/abs/0810.3159} {arXiv:0810.3159 [hep-th]} \BibitemShut
  {NoStop}%
\bibitem [{\citenamefont {Gaiotto}\ and\ \citenamefont
  {Witten}(2009)}]{Gaiotto:2008ak}%
  \BibitemOpen
  \bibfield  {author} {\bibinfo {author} {\bibfnamefont {D.}~\bibnamefont
  {Gaiotto}}\ and\ \bibinfo {author} {\bibfnamefont {E.}~\bibnamefont
  {Witten}},\ }\bibfield  {title} {\bibinfo {title} {{S-Duality of Boundary
  Conditions In N=4 Super Yang-Mills Theory}},\ }\href
  {https://doi.org/10.4310/ATMP.2009.v13.n3.a5} {\bibfield  {journal} {\bibinfo
   {journal} {Adv. Theor. Math. Phys.}\ }\textbf {\bibinfo {volume} {13}},\
  \bibinfo {pages} {721} (\bibinfo {year} {2009})},\ \Eprint
  {https://arxiv.org/abs/0807.3720} {arXiv:0807.3720 [hep-th]} \BibitemShut
  {NoStop}%
\bibitem [{\citenamefont {Kapustin}\ and\ \citenamefont
  {Tikhonov}(2009)}]{Kapustin:2009av}%
  \BibitemOpen
  \bibfield  {author} {\bibinfo {author} {\bibfnamefont {A.}~\bibnamefont
  {Kapustin}}\ and\ \bibinfo {author} {\bibfnamefont {M.}~\bibnamefont
  {Tikhonov}},\ }\bibfield  {title} {\bibinfo {title} {{Abelian duality, walls
  and boundary conditions in diverse dimensions}},\ }\href
  {https://doi.org/10.1088/1126-6708/2009/11/006} {\bibfield  {journal}
  {\bibinfo  {journal} {JHEP}\ }\textbf {\bibinfo {volume} {11}},\ \bibinfo
  {pages} {006}},\ \Eprint {https://arxiv.org/abs/0904.0840} {arXiv:0904.0840
  [hep-th]} \BibitemShut {NoStop}%
\bibitem [{\citenamefont {Weinstein}(1981)}]{weinstein1981symplectic}%
  \BibitemOpen
  \bibfield  {author} {\bibinfo {author} {\bibfnamefont {A.}~\bibnamefont
  {Weinstein}},\ }\bibfield  {title} {\bibinfo {title} {Symplectic geometry},\
  }\href@noop {} {\bibfield  {journal} {\bibinfo  {journal} {Bulletin of the
  American mathematical society}\ }\textbf {\bibinfo {volume} {5}},\ \bibinfo
  {pages} {1} (\bibinfo {year} {1981})}\BibitemShut {NoStop}%
\bibitem [{\citenamefont {Weinstein}(1982)}]{weinstein1982symplectic}%
  \BibitemOpen
  \bibfield  {author} {\bibinfo {author} {\bibfnamefont {A.}~\bibnamefont
  {Weinstein}},\ }\bibfield  {title} {\bibinfo {title} {The symplectic
  “category”},\ }in\ \href@noop {} {\emph {\bibinfo {booktitle}
  {Differential geometric methods in mathematical physics}}}\ (\bibinfo
  {publisher} {Springer},\ \bibinfo {year} {1982})\ pp.\ \bibinfo {pages}
  {45--51}\BibitemShut {NoStop}%
\bibitem [{\citenamefont {Wu}\ and\ \citenamefont {Yang}(1976)}]{wu1976dirac}%
  \BibitemOpen
  \bibfield  {author} {\bibinfo {author} {\bibfnamefont {T.~T.}\ \bibnamefont
  {Wu}}\ and\ \bibinfo {author} {\bibfnamefont {C.~N.}\ \bibnamefont {Yang}},\
  }\bibfield  {title} {\bibinfo {title} {Dirac monopole without strings:
  monopole harmonics},\ }\href@noop {} {\bibfield  {journal} {\bibinfo
  {journal} {Nuclear Physics B}\ }\textbf {\bibinfo {volume} {107}},\ \bibinfo
  {pages} {365} (\bibinfo {year} {1976})}\BibitemShut {NoStop}%
\bibitem [{\citenamefont {Alvarez}(1985)}]{alvarez1985topological}%
  \BibitemOpen
  \bibfield  {author} {\bibinfo {author} {\bibfnamefont {O.}~\bibnamefont
  {Alvarez}},\ }\bibfield  {title} {\bibinfo {title} {Topological quantization
  and cohomology},\ }\href@noop {} {\bibfield  {journal} {\bibinfo  {journal}
  {Communications in Mathematical Physics}\ }\textbf {\bibinfo {volume}
  {100}},\ \bibinfo {pages} {279} (\bibinfo {year} {1985})}\BibitemShut
  {NoStop}%
\bibitem [{\citenamefont {Wehrheim}\ and\ \citenamefont
  {Woodward}(2010)}]{wehrheim2010quilted}%
  \BibitemOpen
  \bibfield  {author} {\bibinfo {author} {\bibfnamefont {K.}~\bibnamefont
  {Wehrheim}}\ and\ \bibinfo {author} {\bibfnamefont {C.~T.}\ \bibnamefont
  {Woodward}},\ }\bibfield  {title} {\bibinfo {title} {Quilted floer
  cohomology},\ }\href@noop {} {\bibfield  {journal} {\bibinfo  {journal}
  {Geometry \& Topology}\ }\textbf {\bibinfo {volume} {14}},\ \bibinfo {pages}
  {833} (\bibinfo {year} {2010})}\BibitemShut {NoStop}%
\bibitem [{\citenamefont {Abouzaid}\ and\ \citenamefont
  {Bottman}(2022)}]{abouzaid2022functoriality}%
  \BibitemOpen
  \bibfield  {author} {\bibinfo {author} {\bibfnamefont {M.}~\bibnamefont
  {Abouzaid}}\ and\ \bibinfo {author} {\bibfnamefont {N.}~\bibnamefont
  {Bottman}},\ }\bibfield  {title} {\bibinfo {title} {Functoriality in
  categorical symplectic geometry},\ }\href@noop {} {\bibfield  {journal}
  {\bibinfo  {journal} {arXiv preprint arXiv:2210.11159}\ } (\bibinfo {year}
  {2022})}\BibitemShut {NoStop}%
\bibitem [{\citenamefont {Wong}\ and\ \citenamefont
  {Affleck}(1994)}]{wong1994tunneling}%
  \BibitemOpen
  \bibfield  {author} {\bibinfo {author} {\bibfnamefont {E.}~\bibnamefont
  {Wong}}\ and\ \bibinfo {author} {\bibfnamefont {I.}~\bibnamefont {Affleck}},\
  }\bibfield  {title} {\bibinfo {title} {Tunneling in quantum wires: A boundary
  conformal field theory approach},\ }\href@noop {} {\bibfield  {journal}
  {\bibinfo  {journal} {Nuclear Physics B}\ }\textbf {\bibinfo {volume}
  {417}},\ \bibinfo {pages} {403} (\bibinfo {year} {1994})}\BibitemShut
  {NoStop}%
\bibitem [{\citenamefont {Lozano}(1995)}]{Lozano:1995aq}%
  \BibitemOpen
  \bibfield  {author} {\bibinfo {author} {\bibfnamefont {Y.}~\bibnamefont
  {Lozano}},\ }\bibfield  {title} {\bibinfo {title} {{S duality in gauge
  theories as a canonical transformation}},\ }\href
  {https://doi.org/10.1016/0370-2693(95)01081-1} {\bibfield  {journal}
  {\bibinfo  {journal} {Phys. Lett. B}\ }\textbf {\bibinfo {volume} {364}},\
  \bibinfo {pages} {19} (\bibinfo {year} {1995})},\ \Eprint
  {https://arxiv.org/abs/hep-th/9508021} {arXiv:hep-th/9508021} \BibitemShut
  {NoStop}%
\bibitem [{\citenamefont {Lozano}(1997)}]{Lozano:1997cy}%
  \BibitemOpen
  \bibfield  {author} {\bibinfo {author} {\bibfnamefont {Y.}~\bibnamefont
  {Lozano}},\ }\bibfield  {title} {\bibinfo {title} {{D-brane dualities as
  canonical transformations}},\ }\href
  {https://doi.org/10.1016/S0370-2693(97)00292-X} {\bibfield  {journal}
  {\bibinfo  {journal} {Phys. Lett. B}\ }\textbf {\bibinfo {volume} {399}},\
  \bibinfo {pages} {233} (\bibinfo {year} {1997})},\ \Eprint
  {https://arxiv.org/abs/hep-th/9701186} {arXiv:hep-th/9701186} \BibitemShut
  {NoStop}%
\bibitem [{\citenamefont {Igarashi}\ \emph
  {et~al.}(1998{\natexlab{a}})\citenamefont {Igarashi}, \citenamefont {Itoh},\
  and\ \citenamefont {Kamimura}}]{igarashi1998self}%
  \BibitemOpen
  \bibfield  {author} {\bibinfo {author} {\bibfnamefont {Y.}~\bibnamefont
  {Igarashi}}, \bibinfo {author} {\bibfnamefont {K.}~\bibnamefont {Itoh}},\
  and\ \bibinfo {author} {\bibfnamefont {K.}~\bibnamefont {Kamimura}},\
  }\bibfield  {title} {\bibinfo {title} {Self-duality in super d3-brane
  action},\ }\href@noop {} {\bibfield  {journal} {\bibinfo  {journal} {Nuclear
  Physics B}\ }\textbf {\bibinfo {volume} {536}},\ \bibinfo {pages} {469}
  (\bibinfo {year} {1998}{\natexlab{a}})}\BibitemShut {NoStop}%
\bibitem [{\citenamefont {Igarashi}\ \emph
  {et~al.}(1998{\natexlab{b}})\citenamefont {Igarashi}, \citenamefont {Itoh},
  \citenamefont {Kamimura},\ and\ \citenamefont
  {Kuriki}}]{igarashi1998canonical}%
  \BibitemOpen
  \bibfield  {author} {\bibinfo {author} {\bibfnamefont {Y.}~\bibnamefont
  {Igarashi}}, \bibinfo {author} {\bibfnamefont {K.}~\bibnamefont {Itoh}},
  \bibinfo {author} {\bibfnamefont {K.}~\bibnamefont {Kamimura}},\ and\
  \bibinfo {author} {\bibfnamefont {R.}~\bibnamefont {Kuriki}},\ }\bibfield
  {title} {\bibinfo {title} {Canonical equivalence between super d-string and
  type iib superstring},\ }\href@noop {} {\bibfield  {journal} {\bibinfo
  {journal} {Journal of High Energy Physics}\ }\textbf {\bibinfo {volume}
  {1998}},\ \bibinfo {pages} {002} (\bibinfo {year}
  {1998}{\natexlab{b}})}\BibitemShut {NoStop}%
\bibitem [{\citenamefont {Alvarez}\ \emph {et~al.}(1994)\citenamefont
  {Alvarez}, \citenamefont {Alvarez-Gaume},\ and\ \citenamefont
  {Lozano}}]{alvarez1994canonical}%
  \BibitemOpen
  \bibfield  {author} {\bibinfo {author} {\bibfnamefont {E.}~\bibnamefont
  {Alvarez}}, \bibinfo {author} {\bibfnamefont {L.}~\bibnamefont
  {Alvarez-Gaume}},\ and\ \bibinfo {author} {\bibfnamefont {Y.}~\bibnamefont
  {Lozano}},\ }\bibfield  {title} {\bibinfo {title} {A canonical approach to
  duality transformations},\ }\href@noop {} {\bibfield  {journal} {\bibinfo
  {journal} {Physics Letters B}\ }\textbf {\bibinfo {volume} {336}},\ \bibinfo
  {pages} {183} (\bibinfo {year} {1994})}\BibitemShut {NoStop}%
\bibitem [{\citenamefont {Klimcik}\ and\ \citenamefont
  {Severa}(1996)}]{Klimcik:1995dy}%
  \BibitemOpen
  \bibfield  {author} {\bibinfo {author} {\bibfnamefont {C.}~\bibnamefont
  {Klimcik}}\ and\ \bibinfo {author} {\bibfnamefont {P.}~\bibnamefont
  {Severa}},\ }\bibfield  {title} {\bibinfo {title} {{Poisson-Lie T duality and
  loop groups of Drinfeld doubles}},\ }\href
  {https://doi.org/10.1016/0370-2693(96)00025-1} {\bibfield  {journal}
  {\bibinfo  {journal} {Phys. Lett. B}\ }\textbf {\bibinfo {volume} {372}},\
  \bibinfo {pages} {65} (\bibinfo {year} {1996})},\ \Eprint
  {https://arxiv.org/abs/hep-th/9512040} {arXiv:hep-th/9512040} \BibitemShut
  {NoStop}%
\bibitem [{\citenamefont {Sfetsos}(1997)}]{Sfetsos:1996xj}%
  \BibitemOpen
  \bibfield  {author} {\bibinfo {author} {\bibfnamefont {K.}~\bibnamefont
  {Sfetsos}},\ }\bibfield  {title} {\bibinfo {title} {{Poisson-Lie T duality
  and supersymmetry}},\ }\href {https://doi.org/10.1016/S0920-5632(97)00339-3}
  {\bibfield  {journal} {\bibinfo  {journal} {Nucl. Phys. B Proc. Suppl.}\
  }\textbf {\bibinfo {volume} {56}},\ \bibinfo {pages} {302} (\bibinfo {year}
  {1997})},\ \Eprint {https://arxiv.org/abs/hep-th/9611199}
  {arXiv:hep-th/9611199} \BibitemShut {NoStop}%
\bibitem [{\citenamefont {Sfetsos}\ \emph {et~al.}(2011)\citenamefont
  {Sfetsos}, \citenamefont {Siampos},\ and\ \citenamefont
  {Thompson}}]{Sfetsos:2010xa}%
  \BibitemOpen
  \bibfield  {author} {\bibinfo {author} {\bibfnamefont {K.}~\bibnamefont
  {Sfetsos}}, \bibinfo {author} {\bibfnamefont {K.}~\bibnamefont {Siampos}},\
  and\ \bibinfo {author} {\bibfnamefont {D.~C.}\ \bibnamefont {Thompson}},\
  }\bibfield  {title} {\bibinfo {title} {{Canonical pure spinor (Fermionic)
  T-duality}},\ }\href {https://doi.org/10.1088/0264-9381/28/5/055010}
  {\bibfield  {journal} {\bibinfo  {journal} {Class. Quant. Grav.}\ }\textbf
  {\bibinfo {volume} {28}},\ \bibinfo {pages} {055010} (\bibinfo {year}
  {2011})},\ \Eprint {https://arxiv.org/abs/1007.5142} {arXiv:1007.5142
  [hep-th]} \BibitemShut {NoStop}%
\bibitem [{\citenamefont {Elitzur}\ \emph {et~al.}(2013)\citenamefont
  {Elitzur}, \citenamefont {Karni}, \citenamefont {Rabinovici},\ and\
  \citenamefont {Sarkissian}}]{Elitzur:2013ut}%
  \BibitemOpen
  \bibfield  {author} {\bibinfo {author} {\bibfnamefont {S.}~\bibnamefont
  {Elitzur}}, \bibinfo {author} {\bibfnamefont {B.}~\bibnamefont {Karni}},
  \bibinfo {author} {\bibfnamefont {E.}~\bibnamefont {Rabinovici}},\ and\
  \bibinfo {author} {\bibfnamefont {G.}~\bibnamefont {Sarkissian}},\ }\bibfield
   {title} {\bibinfo {title} {{Defects, Super-Poincar\'e line bundle and
  Fermionic T-duality}},\ }\href {https://doi.org/10.1007/JHEP04(2013)088}
  {\bibfield  {journal} {\bibinfo  {journal} {JHEP}\ }\textbf {\bibinfo
  {volume} {04}},\ \bibinfo {pages} {088}},\ \Eprint
  {https://arxiv.org/abs/1301.6639} {arXiv:1301.6639 [hep-th]} \BibitemShut
  {NoStop}%
\bibitem [{\citenamefont {Demulder}\ and\ \citenamefont
  {Raml}(2022)}]{Demulder:2022nlz}%
  \BibitemOpen
  \bibfield  {author} {\bibinfo {author} {\bibfnamefont {S.}~\bibnamefont
  {Demulder}}\ and\ \bibinfo {author} {\bibfnamefont {T.}~\bibnamefont
  {Raml}},\ }\bibfield  {title} {\bibinfo {title} {{Poisson-Lie T-duality
  defects and target space fusion}},\ }\href@noop {} {\  (\bibinfo {year}
  {2022})},\ \Eprint {https://arxiv.org/abs/2208.04662} {arXiv:2208.04662
  [hep-th]} \BibitemShut {NoStop}%
\bibitem [{\citenamefont {Arvanitakis}\ \emph {et~al.}(2021)\citenamefont
  {Arvanitakis}, \citenamefont {Blair},\ and\ \citenamefont
  {Thompson}}]{Arvanitakis:2021lwo}%
  \BibitemOpen
  \bibfield  {author} {\bibinfo {author} {\bibfnamefont {A.~S.}\ \bibnamefont
  {Arvanitakis}}, \bibinfo {author} {\bibfnamefont {C.~D.~A.}\ \bibnamefont
  {Blair}},\ and\ \bibinfo {author} {\bibfnamefont {D.~C.}\ \bibnamefont
  {Thompson}},\ }\bibfield  {title} {\bibinfo {title} {{A QP perspective on
  topology change in Poisson-Lie T-duality}},\ }\href@noop {} {\  (\bibinfo
  {year} {2021})},\ \Eprint {https://arxiv.org/abs/2110.08179}
  {arXiv:2110.08179 [hep-th]} \BibitemShut {NoStop}%
\bibitem [{\citenamefont {Bouwknegt}\ \emph
  {et~al.}(2004{\natexlab{a}})\citenamefont {Bouwknegt}, \citenamefont
  {Evslin},\ and\ \citenamefont {Mathai}}]{bouwknegt2004t}%
  \BibitemOpen
  \bibfield  {author} {\bibinfo {author} {\bibfnamefont {P.}~\bibnamefont
  {Bouwknegt}}, \bibinfo {author} {\bibfnamefont {J.}~\bibnamefont {Evslin}},\
  and\ \bibinfo {author} {\bibfnamefont {V.}~\bibnamefont {Mathai}},\
  }\bibfield  {title} {\bibinfo {title} {T-duality: topology change from
  h-flux},\ }\href@noop {} {\bibfield  {journal} {\bibinfo  {journal}
  {Communications in mathematical physics}\ }\textbf {\bibinfo {volume}
  {249}},\ \bibinfo {pages} {383} (\bibinfo {year}
  {2004}{\natexlab{a}})}\BibitemShut {NoStop}%
\bibitem [{\citenamefont {Bouwknegt}\ \emph
  {et~al.}(2004{\natexlab{b}})\citenamefont {Bouwknegt}, \citenamefont
  {Hannabuss},\ and\ \citenamefont {Mathai}}]{bouwknegt2004s}%
  \BibitemOpen
  \bibfield  {author} {\bibinfo {author} {\bibfnamefont {P.}~\bibnamefont
  {Bouwknegt}}, \bibinfo {author} {\bibfnamefont {K.}~\bibnamefont
  {Hannabuss}},\ and\ \bibinfo {author} {\bibfnamefont {V.}~\bibnamefont
  {Mathai}},\ }\bibfield  {title} {\bibinfo {title} {T-duality for principal
  torus bundles},\ }\href@noop {} {\bibfield  {journal} {\bibinfo  {journal}
  {Journal of High Energy Physics}\ }\textbf {\bibinfo {volume} {2004}},\
  \bibinfo {pages} {018} (\bibinfo {year} {2004}{\natexlab{b}})}\BibitemShut
  {NoStop}%
\bibitem [{\citenamefont {Ikehashi}(1993)}]{Ikehashi:1993np}%
  \BibitemOpen
  \bibfield  {author} {\bibinfo {author} {\bibfnamefont {T.}~\bibnamefont
  {Ikehashi}},\ }\bibfield  {title} {\bibinfo {title} {{Hamiltonian formulation
  of the smooth bosonization and local gauge symmetry of the massless Thirring
  model}},\ }\href {https://doi.org/10.1016/0370-2693(93)91197-U} {\bibfield
  {journal} {\bibinfo  {journal} {Phys. Lett. B}\ }\textbf {\bibinfo {volume}
  {313}},\ \bibinfo {pages} {103} (\bibinfo {year} {1993})},\ \Eprint
  {https://arxiv.org/abs/hep-th/9307136} {arXiv:hep-th/9307136} \BibitemShut
  {NoStop}%
\bibitem [{\citenamefont {Bordner}(1997)}]{Bordner:1997pv}%
  \BibitemOpen
  \bibfield  {author} {\bibinfo {author} {\bibfnamefont {A.~J.}\ \bibnamefont
  {Bordner}},\ }\bibfield  {title} {\bibinfo {title} {{Smooth bosonization as a
  quantum canonical transformation}},\ }\href
  {https://doi.org/10.1103/PhysRevD.55.7739} {\bibfield  {journal} {\bibinfo
  {journal} {Phys. Rev. D}\ }\textbf {\bibinfo {volume} {55}},\ \bibinfo
  {pages} {7739} (\bibinfo {year} {1997})},\ \Eprint
  {https://arxiv.org/abs/hep-th/9702193} {arXiv:hep-th/9702193} \BibitemShut
  {NoStop}%
\bibitem [{\citenamefont {Wehrheim}\ and\ \citenamefont
  {Woodward}(2011)}]{wehrheim2011quilted}%
  \BibitemOpen
  \bibfield  {author} {\bibinfo {author} {\bibfnamefont {K.}~\bibnamefont
  {Wehrheim}}\ and\ \bibinfo {author} {\bibfnamefont {C.~T.}\ \bibnamefont
  {Woodward}},\ }\bibfield  {title} {\bibinfo {title} {Quilted floer
  trajectories with constant components},\ }\href@noop {} {\bibfield  {journal}
  {\bibinfo  {journal} {arXiv preprint arXiv:1101.3770}\ } (\bibinfo {year}
  {2011})}\BibitemShut {NoStop}%
\bibitem [{\citenamefont {Weinstein}(2010)}]{weinstein2010symplectic}%
  \BibitemOpen
  \bibfield  {author} {\bibinfo {author} {\bibfnamefont {A.}~\bibnamefont
  {Weinstein}},\ }\bibfield  {title} {\bibinfo {title} {Symplectic
  categories},\ }\href@noop {} {\bibfield  {journal} {\bibinfo  {journal}
  {Portugaliae Mathematica}\ }\textbf {\bibinfo {volume} {67}},\ \bibinfo
  {pages} {261} (\bibinfo {year} {2010})}\BibitemShut {NoStop}%
\bibitem [{\citenamefont {(https://mathoverflow.net/users/10839/jonny
  evans)}()}]{436905}%
  \BibitemOpen
  \bibfield  {author} {\bibinfo {author} {\bibfnamefont {J.~E.}\ \bibnamefont
  {(https://mathoverflow.net/users/10839/jonny evans)}},\ }\href
  {https://mathoverflow.net/q/436905} {\bibinfo {title} {Progress on
  composition of lagrangian correspondences/definition of symplectic
  categories?}},\ \bibinfo {howpublished} {MathOverflow},\ \bibinfo {note}
  {uRL:https://mathoverflow.net/q/436905 (version: 2022-12-20)},\ \Eprint
  {https://arxiv.org/abs/https://mathoverflow.net/q/436905}
  {https://mathoverflow.net/q/436905} \BibitemShut {NoStop}%
\end{thebibliography}%
\end{document}